----------------------------------------------------------------------------
\def\annp #1 #2 #3 {{\sl Ann.\ Phys.\ (N.Y.)} {\bf #1}, #2 (#3)}
\def\jpa #1 #2 #3 {{\sl J. Phys.\ A} {\bf #1}, #2 (#3)}
\def\pra #1 #2 #3 {{\sl Phys.\ Rev.\ A} {\bf #1}, #2 (#3)}
\def\pre #1 #2 #3 {{\sl Phys.\ Rev.\ E} {\bf #1}, #2 (#3)}
\def\prl #1 #2 #3 {{\sl Phys.\ Rev.\ Lett.} {\bf #1}, #2 (#3)}
\def\jsp #1 #2 #3 {{\sl J. Stat.\ Phys.} {\bf #1}, #2 (#3)}
\def\zpb #1 #2 #3 {{\sl Z. Phys.\ B} {\bf #1}, #2 (#3)}
\def\jcp #1 #2 #3 {{\sl J.\ Chem.\ Phys.} {\bf #1}, #2 (#3)}
\def\anp #1 #2 #3 {{\sl Ann.\ Prob.\ } {\bf #1}, #2 (#3)}
\def\annp #1 #2 #3 {{\sl Ann.\ Phys.\ (N.Y.)} {\bf #1}, #2 (#3)}

\def\vs{{\it vs.}}

\def\p2d#1#2{{\partial^2 #1\over\partial #2^2}} 
\def\gtwid{\mathrel{\raise.3ex\hbox{$>$\kern-.75em\lower1ex\hbox{$\sim$}}}}
\def\ltwid{\mathrel{\raise.3ex\hbox{$<$\kern-.75em\lower1ex\hbox{$\sim$}}}}
\def\l{\lambda}
\def\t{\theta}
\def\r{\rho}
\def\en{E_n}
\def\ent{E_n(t)}
\def\rn{R_n}
\def\rnt{R_n(t)}

\documentstyle[12pt]{article}
\begin{document}

\begin{titlepage}
\centerline{\bf Cluster Approximation for the Contact Process}
\bigskip
\centerline{\bf E.~Ben-Naim and P.~L.~Krapivsky}\smallskip
\centerline{Center for Polymer Studies and Department of Physics}
\centerline{Boston University, Boston, MA 02215 }
\vskip .65in
\centerline{ABSTRACT}
{\narrower\narrower\noindent The one-dimensional contact process is
analyzed by a cluster approximation.  In this approach, the hierarchy of
rate equations for the densities of finite length empty intervals are
truncated under the assumption that adjacent intervals are not
correlated.  This assumption yields a first order phase transition from
an active state to the adsorbing state.  Despite the apparent failure of
this approximation in describing the critical behavior, our approach
provides an accurate description of the steady state properties for a
significant range of desorption rates.  Moreover, the resulting critical
exponents are closer to the simulation values in comparison with site
mean-field theory.}
\end{titlepage}

The contact process (CP) is an irreversible lattice model involving
nearest neighbor interactions only \cite{Harris-74,Ligget-85}.  This
model incorporates spontaneous desorption and nearest-neighbor induced
adsorption.  This stochastic process can be used to mimic epidemic
spread as well as catalytic reactions.  This model belongs to a general
class of nonequilibrium models exhibiting a continuous phase transition.
Near the critical point, the system exhibits divergence of spatial and
temporal correlations.  Such properties, conveniently characterized by
critical exponents, can be used to classify different models.  The CP
belongs to the same universality class as Scl\"ogel's first
model \cite{Schlogl-72}, Reggeon field theory
\cite{Cardy-80}, directed percolation \cite{Kinzel-85}, and the ZGB
model \cite{Ziff-86} of catalysis.  Field theoretic renormalization
group studies
\cite{Janssen-81, Grassberger-79}
provide considerable understanding of the critical behavior of the CP.
However, the best estimates for
the characteristic exponents were found numerically by Monte Carlo
simulations \cite{Jensen-92} and by series expansion analysis
\cite{Dickman-91,Jensen-93} .

Motivated by the incomplete theoretical understanding,
we introduce an approximate approach to the CP.  We
study the temporal evolution of the density of empty intervals.  The
corresponding rate equations lead to an infinite hierarchy of
equations. By writing the density of pairs of neighboring empty
intervals as a product over single interval densities, we obtain a
closed set of equations. We use the generating function technique to
obtain the steady-state properties of the system.  Within this
approximation, the system exhibits a discontinuous phase transition from
an active state to the empty state.  As the system approaches the
critical point, the relaxation time, associated with the temporal
approach to the final state, diverges.  Consequently, at the critical
point, an anomalously slow decay towards the final state takes place.
We find the corresponding kinetic exponent by scaling techniques, as
well as by numeric integration of the rate equations.

We compare the cluster approximation predictions with the results of
site mean-field theory and with series analysis of this process.  Despite the
failure to predict a continuous transition, the cluster approximation
provides a good approximation for the final density and the empty interval
density for a reasonable range of desorption rates.  Moreover, the
resulting estimates for the critical exponents are closer to the numeric
values in comparison with site mean-field theory.  The cluster approach
is also applicable to generalizations of the contact process, such as
the A model and the N3 model. We verify that the resulting critical
behavior of these processes is identical with the contact process.
Our approach is advantageous since it can be improved systematically by
considering the evolution of higher order empty interval densities.

In the CP, a particle desorbs spontaneously with rate $\l$. On the other
hand, a particle adsorbs at a given site at a rate proportional to the
number of neighboring occupied sites.  In other words, the adsorption
rate at a particular site is given by $n_p/n_s$, with $n_s$ the total
number of neighboring sites and $n_p$ the number of neighboring
particles. Since every neighboring particle contributes independently to
the adsorption rate, this stochastic process can viewed as an
interacting particle system with nearest-neighbor interactions only. The
above process possesses an adsorbing ``vacuum'' state: once the system
reaches the empty state, adsorption becomes impossible.  In sufficiently
high dimensions, neighboring sites are not correlated, and the density
follows from $d\r/dt=\r(1-\r)-\l\r$.  The adsorption term represents
the density of vacant sites that neighbor an occupied site.
This site mean-field theory (SMF)
gives a steady state concentration equal to $1-\l$. Hence, at $\l_c=1$
this process undergoes a simple continuous transition.

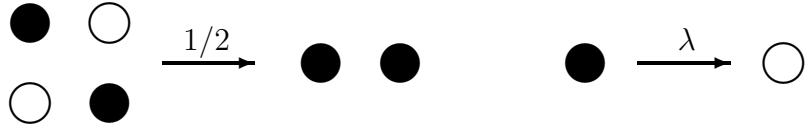
\begin{figure}[t]
\centerline{\begin{picture}(303,48)(6,719)
\thicklines
\put( 15,758){\circle*{15}}
\put( 45,758){\circle{15}}
\put( 15,728){\circle{15}}
\put( 45,728){\circle*{15}}
\put(125,743){\circle*{15}}
\put(155,743){\circle*{15}}
\put(300,743){\circle{15}}
\put(225,743){\circle*{15}}
\put( 65,743){\vector( 1, 0){ 35}}
\put(245,743){\vector( 1, 0){ 35}}
\put( 73,748){1/2}
\put(260,748){$\lambda$}
\end{picture}}
\caption[The Contact Process]{The Contact Process.}
\label{Fig-1}
\end{figure}

We consider the one-dimensional case only (see Figure \ref{Fig-1}),
where studying the density of empty interval has proven useful
in adsorption processes \cite{Cohen-63}, as well as in reaction processes
\cite{Ben-Avraham-90}. Denote by $\ent$ the probability that a randomly
chosen string of $n$ sites is empty (see Figure \ref{Fig-2}).
We emphasize the fact that the actual empty string might be of length
larger than $n$.  Let us also consider $\rnt$, the probability that a
random string of length $n+1$ has $n$ consecutive vacant sites and a
particle at the extreme right site.  For symmetric processes, such as
the CP, $\rnt$ also represent the probability of finding an
empty string with a particle at the extreme left.
These two interval densities are related by
\begin{equation}
R_n=E_n-E_{n+1}\qquad{\rm or}\qquad E_{n}=1-R_0\cdots-R_{n-1},
\label{def}
\end{equation}
for $n\ge0$. For $n=0$, the definition of $E_n$ is trivial,
leading to the following conditions satisfied by the empty interval
densities,
\begin{equation}
\qquad E_0=1\qquad{\rm or}\qquad\sum_{n=0}^{\infty}R_n=1.
\label{cond}
\end{equation}
The condition for $R_n$ is obtained by using Eq.~(\ref{def}) and noting
that the sum over $E_n$ reduces to an alternating series.  The above
interval densities are useful in describing macroscopic properties;
for example, the concentration is given by $\rho(t)=R_0(t)$.
\begin{figure}[t]
\begin{picture}(480,23)(10,749)
\thicklines
\put( 105,758){\circle{12}}
\put(425,758){\circle*{12}}
\put(191,758){\circle{12}}
\put(330,758){\circle{12}}
\put(405,758){\circle{12}}
\put(133,758){\circle*{2}}
\put(148,758){\circle*{2}}
\put(163,758){\circle*{2}}
\put(357,758){\circle*{2}}
\put(368,758){\circle*{2}}
\put(380,758){\circle*{2}}
\put( 14,755){\Large $E_n={\rm Prob}($}
\put(202,755){\Large $)$}
\put(235,755){\Large $R_n={\rm Prob}($}
\put(435,755){\Large $)$}
\put(100,752){\Large$\underbrace{\makebox[97pt]{}}_{n}$}
\put(325,752){\Large$\underbrace{\makebox[85pt]{}}_{n}$}
\end{picture}
\caption[Empty Interval Densities]{Empty Interval Densities}\label{Fig-2}
\end{figure}

We write the rate equations governing $\ent$ by considering the
adsorption and the desorption processes separately.  Adsorption can
contribute only to loss of empty intervals.  Empty intervals of length
$n$ can be destroyed when a particle is desorbed at the edge of the
interval. This occured only where the empty interval has an occupied site
at its edge.  By taking into account contributions from adsorption at
either boundary, we find $d\en/dt\big|_{\rm ads}=-\rn$.  Desorption, on
the other hand, leads only to creation of empty intervals.  When a
particle desorbs, the two empty intervals it borders create a larger
empty interval.  We define the pair density $E_{l,m}$ as the density of
two neighboring empty intervals, with lengths at least equal to $l$ and
$m$, separated by a single particle.
Note that from this definition the relation $E_{0,n}=E_{n,0}=R_n$
is satisfied. In terms of this pair density, the
increase in the density of empty intervals of length at least $n$ due
to desorption is described by
\hbox{$d\en/dt\big|_{\rm des}=\l\sum_{l=0}^{n-1}E_{l,n-1-l}$}.
We then approximate the pair density, $E_{l,m}$, by the product
$E_{l,m}\cong R_lR_m/R_0$, where the factor $1/R_0$ ensures the
normalization condition $E_{0,n}=R_n$.
Combining contributions from both adsorption and desorption
yields the following rate equation for the empty interval density,
\begin{equation}
{d\en \over dt} =-\rn+{\l\over R_0}
{\sum_{l=0}^{n-1}R_lR_{n-1-l}},
\qquad n>0.
\label{rate-eq}
\end{equation}

The steady-state properties can be obtained by requiring that the time
derivative in Eq.~(\ref{rate-eq}) vanishes. We introduce the generating
function $R(z)=\sum_n\rn z^n/R_0$, with $\rn$ being the steady-state interval
densities. Dividing Eq.~(\ref{rate-eq}) by $R_0$, summing over all $n$,
and solving the resulting quadratic equation yields
\begin{equation}
R(z)={1-\sqrt{1-4\l z}\over 2\l z},\qquad {\l<1/4}.
\label{rz}\end{equation}
The normalization condition of Eq.~(\ref{cond}) shows that
$\rho=R_0=1/R(z)\big|_{z=1}$ and, consequently,
the concentration is given by
\begin{equation}
\rho={1+\sqrt{1-4\l}\over 2},\qquad \l<1/4.
\label{rho}
\end{equation}

For $\l>\l_c=1/4$ the system indeed exhibits a transition to the
absorbing state, where the concentration vanishes.
The nature of this transition is discontinuous,
while for the actual CP, the transition is continuous.
For the CP, the approach to the critical density is an algebraic one,
\begin{equation}
\rho-\rho_c\sim(\l_c-\l)^{\beta}.
\label{beta-def}
\end{equation}
Note that while for a continuous transition to a vacuum state one has
$\rho_c=0$, the critical density of the cluster approximation is finite,
$\rho_c=1/2$.  Extensive power series studies suggest $\beta\cong0.277$,
and $\l_c\cong0.3032$ \cite{Dickman-91,Jensen-93}.  The corresponding
values obtained by the cluster approximation, $\beta=1/2$ and
$\l_c=1/4$, are closer than the SMF values $\beta=\l_c=1$. While
the critical point depends heavily on the microscopic definition of the
process, the critical exponents are universal. Applying the cluster
approximation to variants of the CP always yields $\beta=1/2$.

\begin{figure}[t]
\vskip2in
\caption[Steady-state density for the CP]
{The cluster approximation density (solid) \vs\ the actual CP
density (dashed). The latter density represent the [12,12] Pad\'e
approximant obtained from the perturbation study of ref
\cite{Jensen-93}.}
\label{Fig-3}
\end{figure}

In Figure \ref{Fig-3}, we plot the cluster approximation density \vs\ the
series study density.
For desorption rates $\ltwid 0.2$ both curves are
practically identical and, for example, at $\l=0.2$ the relative
difference is less than $0.5\%$. We conclude that despite the failure
near the transition point, the cluster approximation is useful in
describing the process for a substantial dynamic range. Another way to
determine the accuracy of the approximation is by expanding
Eq.~(\ref{rho}) as a power series in $\l^n$, and comparing to the
coefficients, obtained by the series expansion study \cite{Jensen-93}.
We find that both expansions are identical to the third order,
\begin{equation}
\rho(\l)=\cases{
1-\l-\l^2-2\l^3-5\l^4+O(\l^5)&{\rm CA},\cr
1-\l-\l^2-2\l^3-4{1\over 2}\l^4+O(\l^5)&{\rm CP}.}
\label{series}
\end{equation}
Since $R_n$ is of order $\l^n$, we expect similar correspondence between
the approximate interval density from our rate equations and the actual
contact process densities. To test this, we have performed a Monte-Carlo
simulation with $10^5$ particles at $\l=0.2$.  Indeed, the average over
200 different realizations yields values for $R_n$ that agree with the
approximate density to within $0.5\%$, for $n=0,1,2,3$. The quality of
the approximation gradually decreases as $n$ increases and for example
for $n=4$ the discrepancy is $4\%$.

The interval density also exhibits an interesting critical behavior.
By expanding the generating function of Eq.~(\ref{rz}), one finds
$R_n=\r\l^n (2n)!/n!(n+1)!$ for $\l\le\l_c$. This expression shows
that for  $\l<\l_c$ the interval density depends exponentially on
the interval length,
\begin{equation}
\rn\sim n^{-3/2}\exp\Bigl(-4\Delta n\Bigr)\qquad \Delta\ll 1,
\label{rn}\end{equation}
where $\Delta=\l_c-\l$.
The Stirling formula was used to obtain the above form of the interval
density. At the critical point, power-law decay is recovered,
$R_n\sim n^{-3/2}$.

These static properties are closely related to the kinetics of the
system. Far from the critical point, the density rapidly relaxes to its
steady-state value from any initial conditions. As the system
approaches the critical point, the relaxation time diverges and a
power-law decay of the concentration takes place.  We thus expect that
for sufficiently large times and close enough to the critical point, the
temporal approach to the final state is given by the scaling form
\cite{Dickman-91,Racz-85}
\begin{equation}
\rho-\rho_c\sim t^{-\delta}\psi
\bigl(\Delta t^{1/\nu}\bigr).
\label{scaling-c}\end{equation}
In other words, the critical exponent $\delta$ characterizes the
critical kinetics, while the exponent $\nu$ characterizes the
sub-critical relaxation time. The above scaling form should match the
steady-state form of Eq.~(\ref{beta-def}) at large times and hence we
conclude that $\psi(x)\sim x^\beta$ for $x \gg 1$.  To cancel the
temporal dependence, the scaling relation $\beta=\delta\nu$ must be
satisfied.  Furthermore, Eq.~(\ref{rn}) indicates that the relaxation
length associated with the steady-state interval density diverges
as $1/\Delta$ as the system approaches the critical point.
Thus, it is natural to assume that $\rnt$ depends on time through a
rescaled size, $n\to nt^{-\alpha}$, as well as a rescaled adsorption
rate. Noting that $R_0=\rho$, we postulate the following scaling
behavior for $\rnt$:
\begin{equation}
\rnt \sim \phi\Bigl(\Delta t^{1/\nu},nt^{-\alpha}\Bigr).
\label{scaling-rn}
\end{equation}
On the other hand, Eq.~(\ref{rn}) indicates that the
steady-state density depends on the size and the rate only through the variable
$n\Delta$. Hence, $\phi(x,y)\sim\tilde\phi(xy)$, and by eliminating
the temporal dependence we find the scaling relation $\alpha\nu=1$.

\begin{figure}[t]
\vskip2in
\caption[The critical approach towards the steady-state]
{The critical approach towards the steady-state. Numeric solution to the
rate equations (dots) for the case $\l=\l_c=1/4$ is plotted. For
comparison a line of slope $-1/3$ is also shown (solid).}
\label{Fig-4}
\end{figure}

Thus far, our scaling analysis involved matching the anticipated kinetic
behavior to the exact steady-state properties. To determine the critical
temporal decay, \hbox{$\rho-\rho_c\sim t^{-\delta}$}, we study the rate
equations at $\l_c=1/4$.  Using the duality relations between $E_n$ and
$R_n$ (see Eq.~(\ref{def})), the rate equations for $\l=\l_c$ can be
rewritten in terms of $R_n$ only,
\begin{equation}
 \sum_{l=0}^{n-1}{dR_l \over dt} ={1\over 4R_0}
\sum_{l=0}^{n-1} R_l R_{n-1-l}-\rn.
\label{rate-eq-asmp}\end{equation}
To analyze this equation by scaling techniques, we match the leading
asymptotic terms in both sides of the above equation.
The left hand side  is governed by $n$ terms of order $R_n$.
Hence, by taking into account the time derivative,
we conclude that the left hand side is proportional to $nR_n/t$.
The right hand side is dominated by the first few terms in the expansion,
namely $l\ll n$ and $n-l\ll n$. Therefore, we approximate the sum by
$2R_n(R_0+R_1+\cdots)/R_0=2R_n/R_0$, using the normalization condition
of Eq.~(\ref{cond}). Finally, we write the resulting expression
$R_n\left(1-1/2R_0(t)\right)$ in terms of the concentration,
\begin{equation}
{n\over t}R_n\propto R_n(\rho-\rho_c).
\end{equation}
We conclude that $nt^{-1}\sim t^{\alpha-1}\sim t^{-\delta}$, or
equivalently $\alpha+\delta=1$. The three scaling relations yield the
following exponents, $\delta=1/3$, $\nu=3/2$ and $\alpha=2/3$.
Numerical integration of the rate equation at $\l=\l_c$ confirms the
scaling prediction for $\delta$ (see Figure \ref{Fig-4}). In Table
\ref{T-1}, we compare the exponents, that result from our
 cluster approximation (CA), with the corresponding series expansion
and the SMF values.
We conclude that the CA exponents provide significant
improvement in comparison with those from SMF.

\begin{table}[t]
\centerline{\begin{tabular}{|l|c|c|c|}\hline
&$\beta$&$\delta$&$\nu$\\ \hline
CP&0.277&0.160&1.735\\
CA&1/2&1/3&3/2\\
SMF&1&1&1\\ \hline
\end{tabular}}
\caption[The critical exponents]
{Static and Critical decay exponents obtained by series studies (CP),
Cluster Approximations (CA) and Site Mean-Field (SMF).}
\label{T-1}
\end{table}

Recently, several variants of the CP where introduced and it was
shown that they belong to the same universality class as the original
model.  In these models, the adsorption process is modified while the
desorption process remains unchanged.  The adsorption rate is set to
$\t/2$ if only one neighboring site is occupied, while the rate remains
unity in the case when both neighboring sites are occupied.  The empty
interval method can be easily generalized to this case and we merely
quote the resulting subcritical steady state density
\begin{equation}
\r={\t+\sqrt{\t^2-4\l(\t-\eta)}\over2(\t-\eta)},
\qquad{\rm with}\qquad\eta={\l(1-\t)(\t-2\l)\over\t^2+(1-\t)(\t-2\l)}.
\label{t}\end{equation}
For the basic CP, $\t=1$, we recover Eq.~(\ref{rho}).
Clearly, the discontinuous nature of the transition
is independent of the microscopic details of the model
and the exponent $\beta=1/2$ is indeed robust.
The critical point, however, depends on $\t$ and can be found
by equating the square root in Eq.~(\ref{t}) to zero.
If the adsorption rates are independent of the number
of neighboring particles, $\theta=2$,
the predicted critical point is $\l_c\cong0.4608$,
while series studies yield $\l_c\cong0.574$ for this so-called A model.
In the case where $\theta=1/2$ (the N3 model),
the critical point is $\l_c=0.1366$,
while series studies yield $\l_c=0.162$.
For the general $\theta$ case, the cluster approach yields
less accurate estimates for the concentration than for basic CP.

In summary, we have presented an approximate approach to the contact
process.  Based on the assumption that neighboring empty intervals are
not correlated, we solved for the steady state properties.
In addition, the kinetic approach towards the steady state
was found by scaling techniques.  The above approximation is valid for a
significant subcritical range.  The cluster approximation predicts a
discontinuous transition but gives improved exponents in comparison
with simple site mean-field theory.

The cluster approximation can be systematically improved by considering
higher order interval densities. Indeed, our preliminary results
indicate that the second order cluster approximation yields a continuous
phase transition. Moreover, the approximation appears to be valid over a
significantly larger range of desorption rates.

\medskip\centerline{\bf Acknowledgments}\smallskip
We are thankful to R.~Dickman and S.~Redner for useful discussions.
We gratefully acknowledge ARO grant \#DAAH04-93-G-0021, NSF grant
\#DMR-9219845, and to the Donors of The Petroleum Research
Fund, administered by the American Chemical Society, for partial support
of this research.


\end{document}